\documentclass{iopart}
\usepackage{graphicx}
\usepackage{epsfig}
\def\lsim{\raise0.3ex\hbox{$<$\kern-0.75em\raise-1.1ex\hbox{$\sim$}}}
\def\gsim{\raise0.3ex\hbox{$>$\kern-0.75em\raise-1.1ex\hbox{$\sim$}}}
\begin{document}

\title[QCD Thermodynamics at Non-zero Baryon Number Density]{Thermodynamic 
Properties of Strongly Interacting Matter at Non-zero Baryon Number Density}

\author{Frithjof Karsch\footnote[3]{karsch@physik.uni-bielefeld.de}
}
\address{Fakult\"at f\"ur Physik, Universit\"at Bielefeld, D-33615
Bielefeld, Germany} 

\begin{abstract}
We present recent lattice results on QCD thermodynamics at non-vanishing
baryon number density obtained from a 6th order Taylor expansion in the 
chemical potential.  Results for bulk thermodynamic observables, in particular
for fluctuations in the baryon number density, are found to be well described
by a hadron resonance gas model at low temperature and an ideal quark gluon gas 
at high temperature. We also analyze the radius of convergence of the Taylor
series and discuss the information it provides on the occurrence of a second
order phase transition point in the QCD phase diagram.
 
\end{abstract}




\section{Introduction} 

Two quite different aspects of the studies of QCD thermodynamics on the lattice
gained most attention in the heavy ion community during recent years -- 
studies of the QCD phase diagram at non-zero baryon chemical potential and 
the analysis of the influence of a thermal heat bath
on basic properties of hadrons, eg. their masses and widths. Here we will
concentrate on the former. For a recent discussion of in-medium properties of 
hadrons we refer to the proceedings of this years Quark Matter \cite{qm04} 
and Lattice \cite{lat04} conferences.  

Lattice calculations at vanishing baryon chemical potential, $\mu_B$,
suggest that for physical values of the quark masses the transition to the 
high temperature phase of QCD is just a rapid crossover rather than a phase
transition which would be signaled by singularities in bulk thermodynamic
observables.
On the other hand many QCD motivated model calculations
suggest that for $\mu_B > 0$ and small values of the temperature the
transition indeed is a first order phase transition. This suggests that 
somewhere in the interior of the phase diagram there exists a second order
phase transition point (chiral critical point) as an end point of a line
of first order transitions. For vanishing light quark masses this point would 
be a tri-critical point \cite{Stephanov} which,  with increasing value of the 
strange quark mass, is expected to move towards lower values
of $\mu_B$ and eventually could reach the $\mu_B=0$ axis.

\begin{figure}
\begin{center}
\hspace*{-2mm}
\epsfig{bbllx=72,bblly=610,bburx=510,bbury=750,
file=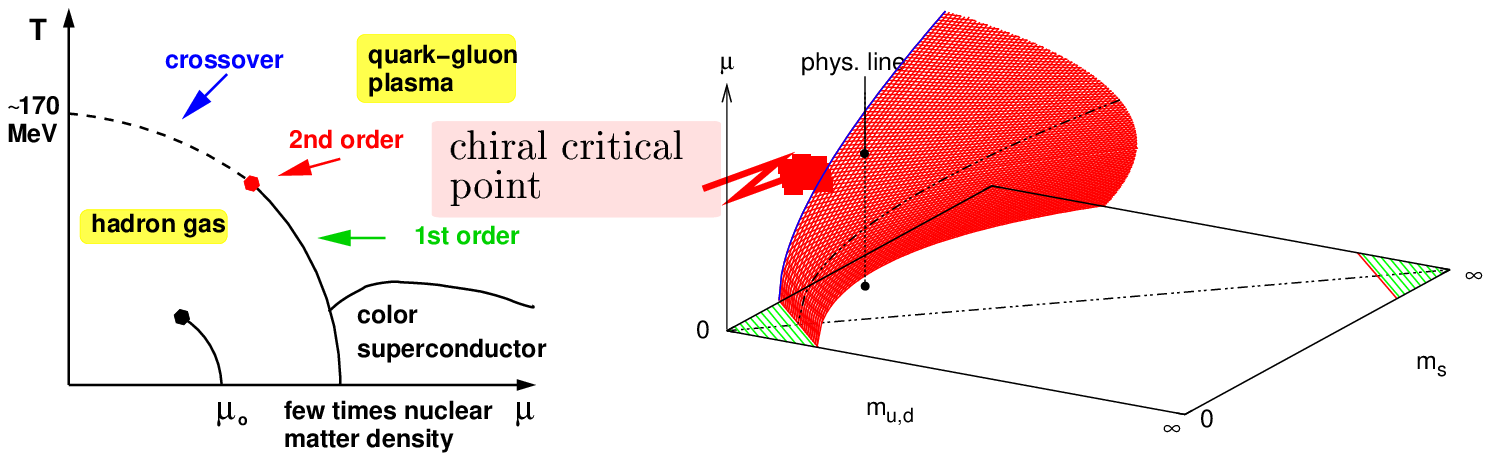,width=13.5cm}
\end{center}
\caption{\label{fig:phased} Generic phase diagram of QCD with physically
realized quark mass values (left). The occurrence of a second order
endpoint in the interior of the phase diagram is expected to be a 
consequence of the existence of a regime of first order transitions 
in 3-flavor QCD at vanishing $\mu_B$ and for small values of the
up, down and strange quark masses. The chiral critical point then lies
on a hyper-surface of second order transition points emerging from the
$\mu_B=0$ plane (right). 
}
\end{figure}

The expected generic phase diagram for physical values of the quark
masses is shown in Fig.~\ref{fig:phased}. Its exploration for large
temperatures and small values of the baryon chemical potential became
accessible to lattice calculations through the application of different
techniques such as Ferrenberg-Swendsen reweighting \cite{Fodor1}, 
Taylor series expansions \cite{Allton1,Gavai1} as well as through
simulations with an imaginary chemical potential \cite{Philipsen1,Lombardo1}. 
This led to a first analysis of
the baryonic contribution to bulk thermodynamic observables like 
the pressure, baryon number density and various susceptibilities
\cite{Fodor2,Allton2,Gavai2,Lombardo}. Moreover, it provided first
estimates for the location of the chiral critical point 
\cite{Fodor1} at which the transition to the high temperature 
phase of QCD turns from a rapid crossover into a first order phase
transition; recent calculations suggest
a critical value $\mu_B^{crit} \sim 400$~MeV \cite{Allton3,Fodor3}.
We will discuss these results in more detail in section 3. However,
before turning to this it seems appropriate to briefly review results on 
the QCD transition at vanishing baryon 
chemical potential where we try to emphasize the role of the strange quark in 
QCD thermodynamics. We will do so in section 2. Section 4 is devoted to 
a comparison of lattice results on thermodynamics in the low
temperature hadronic phase at $\mu_B \;>\; 0$ with a phenomenological
approach in terms of a hadron resonance gas model. This has been discussed
in more detail at this conference by K. Redlich \cite{redlich_sqm}.

\section{The thermal QCD transition at vanishing baryon chemical potential}

For $\mu_B = 0$ the transition from the
low temperature hadronic phase to the high temperature phase has been 
analyzed in many numerical calculations. In particular, it has been found
that for the case of QCD with three degenerate light quarks the transition 
is first order. In fact, the use of improved discretization schemes
for the fermionic part of the QCD Lagrangian has shifted this regime to rather
small values of the quark masses which correspond to a pion mass even 
below its physically realized value \cite{nf3trans}. This strongly 
suggests that the transition is continuous for the physically
realized spectrum of two light ($u, d$)-quarks and a heavier $s$-quark.
A second order phase transition can then only show up at non-vanishing 
chemical potential if the hyper-surface of $2^{nd}$-order phase 
transition, in which also the line of $2^{nd}$-order transitions found
at $\mu_B=0$ in the ($ud, s$)-plane lies, bends over the physical point.
This situation is illustrated in Fig.~\ref{fig:phased}(right). The 
left hand part of this figure shows the resulting phase diagram of QCD with
the physically realized mass spectrum. 

\begin{figure}
\begin{center}
\epsfig{file=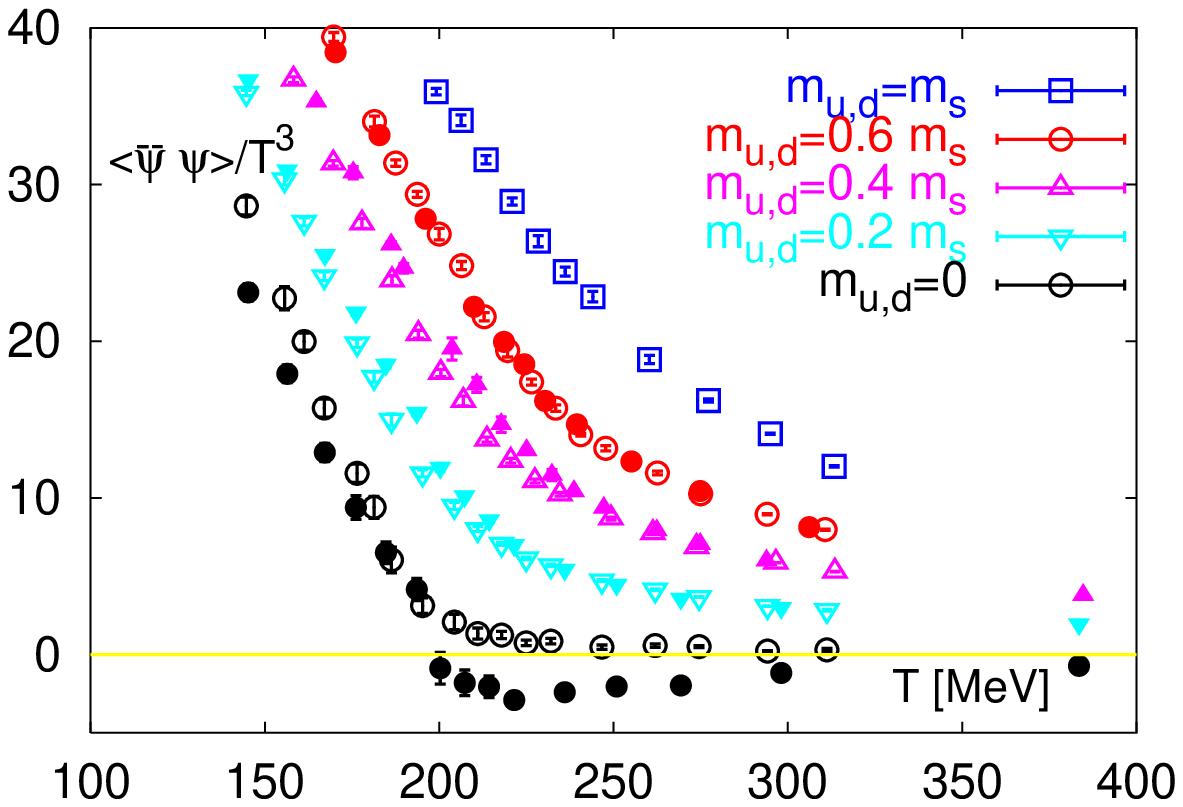,width=64mm}
\epsfig{file=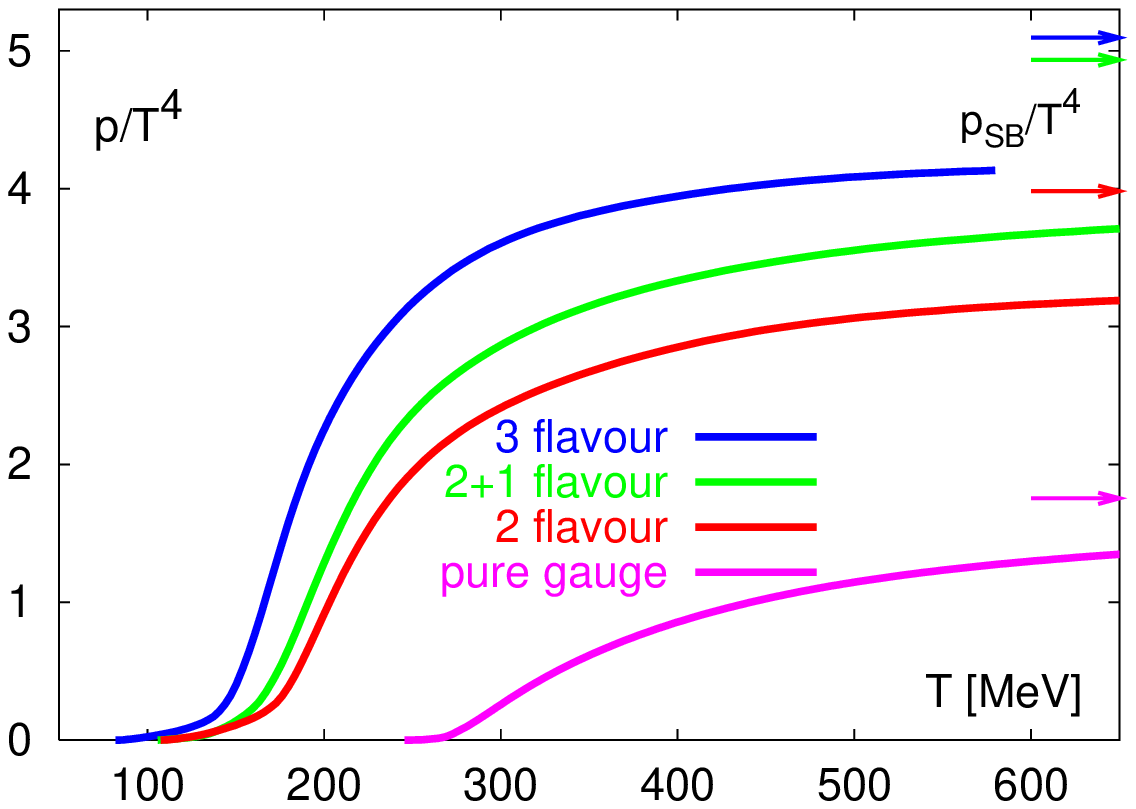,width=64mm}
\end{center}
\caption{\label{fig:chiral} The light quark chiral condensate in QCD
with 2 light up, down and a heavier strange quark mass (open symbols)
and in 3-flavor QCD with degenerate quark masses
(full symbols) \cite{Bernard}. The right hand part of the figure shows the 
pressure calculated in QCD with different number of flavors as well as
in a pure gauge theory \cite{Peikert}. Note that (2+1)-flavor QCD here
refers to QCD with two light quarks and a heavier (strange) quark with a mass 
proportional to the temperature, $m_s\sim T$.
}
\end{figure}

The transition to the high temperature phase of QCD commonly is related
to the restoration of chiral symmetry as well as deconfinement, {\it i.e.}
the disappearance of an almost massless pion as a Goldstone particle and
the sudden liberation of quark and gluon degrees of freedom. This is,
indeed, reflected in the temperature dependence of the chiral condensate
\cite{Bernard} and the the pressure \cite{Peikert} shown in 
Fig.~\ref{fig:chiral}. These figures also indicate that
the presence of strange quarks with a mass $m_s\sim {\cal O}(T)$ has 
little influence on the thermodynamics
in the low temperature phase. The temperature and quark mass dependence
of the light quark chiral condensate in QCD with two light and a heavier 
strange quark is similar to that of QCD with 3 degenerate (light) quarks.
At $T_c$, however, the strange quarks with a mass
$m_s\sim T_c$ do not yet contribute significantly to the pressure or 
energy density.
Consequently the transition temperature in 2 and (2+1)-flavor QCD has been
found to be quite similar. 
In the high temperature phase, on the other hand, bulk thermodynamic
observables clearly reflect the presence of the additional strange
quark degrees of freedom\footnote{In the calculation of the pressure
in (2+1)-flavor QCD shown in Fig.~\ref{fig:chiral} the strange quark mass
has been taken to be proportional to the temperature, {\it i.e.}
$m_s/T=$const. If one keeps instead $m_s$ fixed to its physical value
the pressure will gradually approach the high temperature limit of 3-flavor
QCD. This can be inferred, for instance, from the calculation performed in
\cite{Fodoreos}.}. 
 
This suggests that at least at $\mu_B=0$ the strange quark 
has little influence on the dynamics of the transition. The 
calculations performed with improved actions \cite{Bernard,Peikert,Hasenfratz} 
indicate that for the physically realized quark mass spectrum the
transition to the high temperature phase is signaled by a rapid but continuous
change in bulk thermodynamic observables.
The transition temperature of (2+1)-flavor QCD found in a recent analysis 
with the Naik
action \cite{Bernard}\footnote{Here we have averaged over the two fits
performed in \cite{Bernard} to extrapolate to the chiral limit and included 
systematic uncertainties arising from these different fits 
in the systematic error.},
\begin{equation}
T_c = (172 \pm 11 \pm 7)~{\rm MeV}\quad ,
\end{equation}
is consistent with earlier findings for 2-flavor QCD based on simulations with 
the p4-action, $T_c = (173 \pm 8 \pm {\rm sys.~err.})~{\rm MeV}$
\cite{Peikert}. 

\section{Taylor expansion of the pressure at non-zero baryon chemical potential}

As mentioned in the introduction different approaches have been followed 
to study the phase structure of QCD at non-vanishing
baryon chemical potential and to determine basic thermodynamic observables. 
All these approaches are currently limited to the
regime of large temperatures and $\mu_q/T \lsim 1$, where $\mu_q = \mu_B/3$
is the quark chemical potential. We will concentrate
here on a discussion of Taylor expansions of the partition function of 
2-flavor QCD around $\mu_B=0$.

\begin{figure}
\begin{center}
\epsfig{file=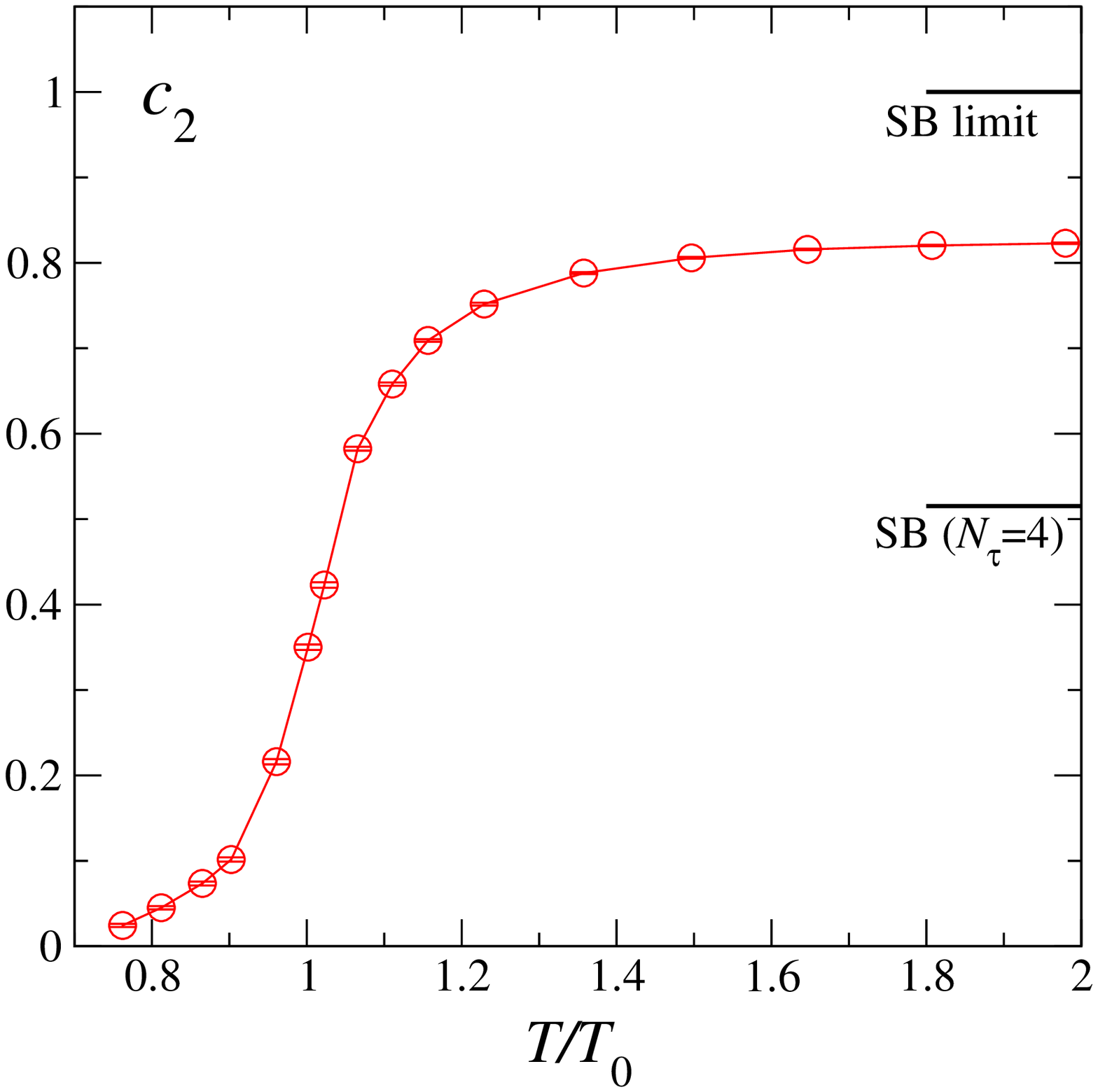,width=4.2cm}
\epsfig{file=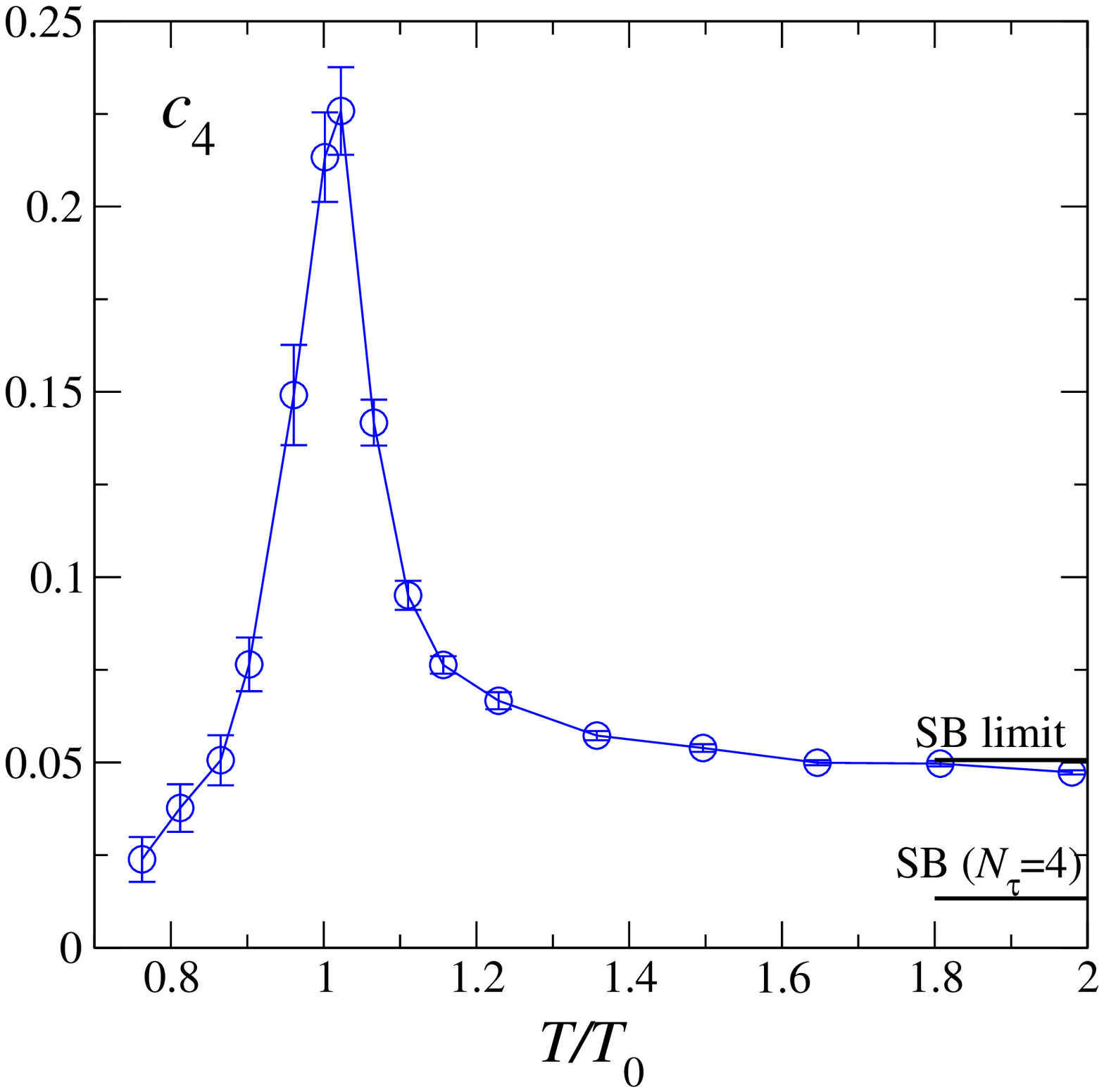,width=4.2cm}
\epsfig{file=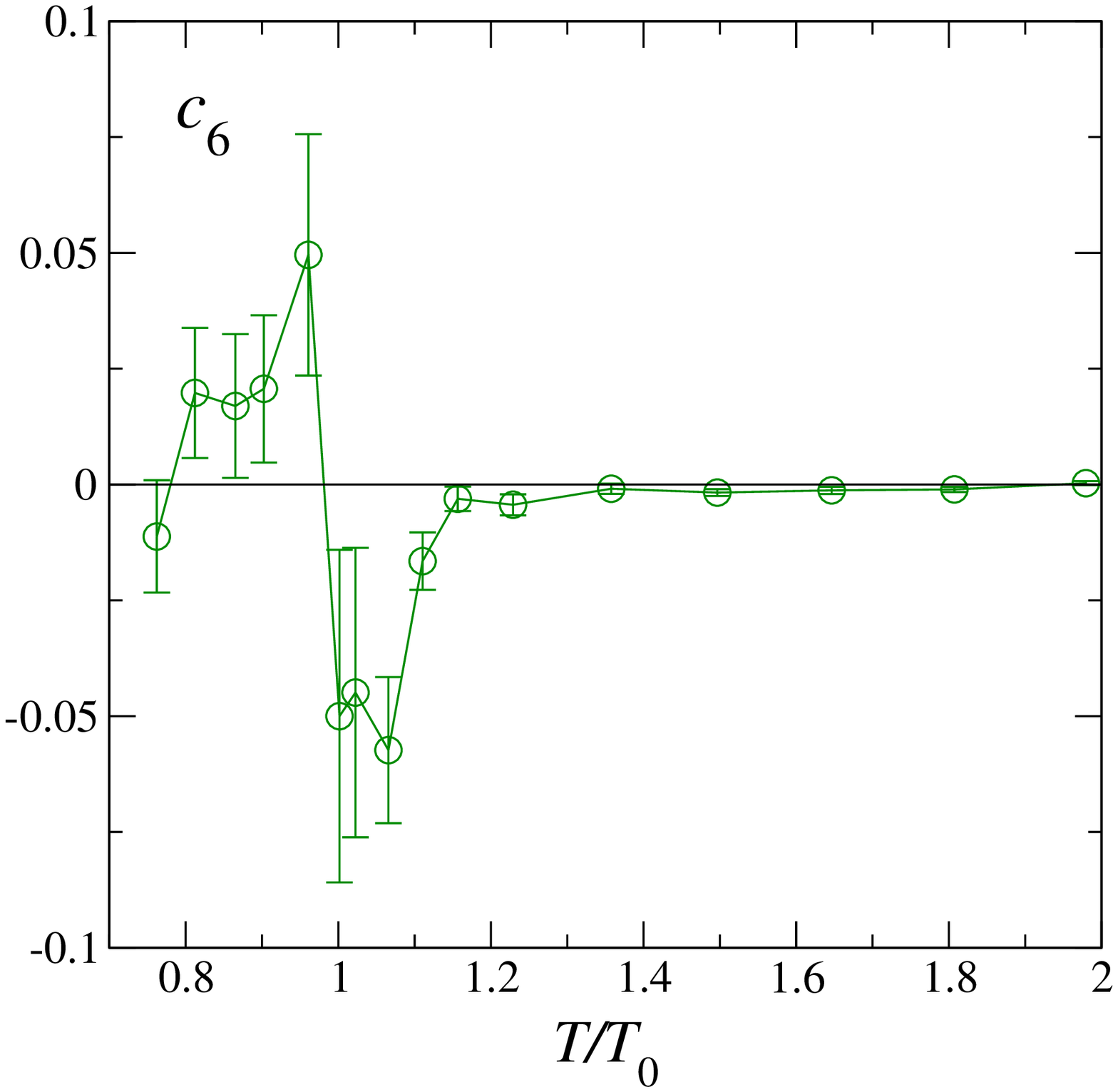,width=4.2cm}
\end{center}
\caption{\label{fig:coefficients} Temperature dependence of expansion 
coefficients for $p/T^4$
in 2-flavor QCD and for quark masses corresponding to a pseudo-scalar (pion)
mass of about $770$~MeV. 
}
\end{figure}

At fixed temperature and small values of the chemical potential the pressure
may be expanded in a Taylor series around $\mu_q = 0$,
\begin{equation}
{p\over T^4}={1\over{VT^3}}\ln{\cal Z} = \sum_{n=0}^{\infty} c_n(T) 
\left( \frac{\mu_q}{T} \right)^n \quad ,
\label{Taylorp}
\end{equation}
where the expansion coefficients are given in terms of derivatives of
$\ln{\cal Z}(V,T,\mu_q)$, {\it i.e.} $c_n(T) = \displaystyle{\frac{1}{n!}
\frac{\partial^n \ln Z}{\partial (\mu_q / T)^n}}$.  The series is 
even in $(\mu_q/T)$ which reflects the invariance of $Z(V,T,\mu_q)$ under
exchange of particles and anti-particles.. The first coefficient, $c_0$,
simply gives the pressure at $\mu_q = 0$ shown in Fig.~\ref{fig:chiral}(right). 
The next non-zero coefficient, $c_2$,
is proportional to the quark number susceptibility at $\mu_q = 0$
\cite{Gottlieb},
\begin{equation}
\frac{\chi_q}{T^2} = \frac{\partial^2 p/T^4}{\partial (\mu_q/T)^2} =
2 c_2 + 12 c_4\; \left( \frac{\mu_q}{T} \right)^2 +{\cal O}\left( (\mu_q/T)^4
\right) \quad .
\label{chiq}
\end{equation}
Its rise with temperature has been attributed to changes in the interaction 
among quarks and anti-quarks in the vector channel \cite{Kunihiro}.
Further Taylor expansion coefficients entering the calculation of the pressure
have been analyzed now up to ${\cal O} (\mu_q^6)$ \cite{Allton2,Gavai3,Allton4}. 
The coefficients for $n=2,$~4 and 6 are shown in Fig.~\ref{fig:coefficients}. 

In the high temperature, ideal gas limit, only the first three expansion 
coefficients are non-zero, 
$c_0(T=\infty) = 7n_{\rm f}\pi^2/60$, $c_2(T=\infty)=n_{\rm f}/2$ 
and $c_4(T=\infty) = n_{\rm f}/4\pi^2$. As can be seen in 
Fig.~\ref{fig:coefficients} at $T\simeq 2 T_0$ all coefficients reach their 
corresponding ideal gas values within $\sim 20\%$. 

The next-to-leading
order coefficient, $c_4$, is strictly positive and develops a pronounced peak
at the $(\mu_q=0)$ transition temperature, $T_0$, which in turn, also leads to
a large peak in the quark number susceptibility 
once $\mu_q/T \gsim 0.5$ \cite{Allton2} (see Fig.~5(right)). 
This has been interpreted as an indication for an approach
to the chiral critical point at which $\chi_q$ is expected to 
diverge\footnote{Fluctuations in the net quark number density are related 
to the isothermal compressibility,
$\kappa_T= -V^{-1}\left( \partial V / \partial p \right)_{T,N_q} = 
-N_q^{-1}\left( \partial V / \partial \mu_q \right)_{T,N_q} = \chi_q/n_q^2$, 
which diverges at a $2^{nd}$ order transition point.}. 

A new feature shows up in the expansion coefficients at ${\cal O}(\mu^6)$.
The coefficient $c_6$ is positive only below $T_0$ and changes sign in its
vicinity. This has important consequences
for the analytic structure of the QCD partition function and gives  
first hints at the convergence properties of the Taylor series: If there
exists a $2^{nd}$ order phase transition point in the QCD phase diagram this
would be related to a Lee-Yang zero of the QCD partition function for
some real and positive value of $(\mu_q/T)^2$. This critical value will be 
directly related to the radius of convergence, $\rho (T)$, of the Taylor 
expansion, if the partition function 
has no further zeroes in the complex ($\mu_q/T$)-plane closer to the origin.
A sufficient condition for the radius of convergence to be due to a zero
on the real ($\mu_q/T$)-axis is that all expansion coefficients in the 
Taylor series are positive. This apparently is the case 
for all $c_n(T)$ analyzed so far at $T\lsim 0.95 T_0$.    

Ratios of subsequent expansion coefficients provide an estimate 
for the radius of convergence of the Taylor series,
\begin{equation}
\rho (T)=\lim_{n\to\infty}\rho_n\equiv\lim_{n\to\infty}\sqrt{\left\vert
{c_n\over c_{n+2}}\right\vert} \quad .
\label{convergence} 
\end{equation}
These ratios are shown in Fig.~\ref{fig:ratios_coefficients}(left).
While the first ratio, $c_4/c_2$ is well determine in current lattice
calculations the ratio $c_6/c_4$ still has large errors. 
It is, however, 
apparent that these ratios change drastically across $T_0$ and seem
to be approximately temperature independent in the low temperature phase.
As we will discuss in the next section this is expected to be the case
also for a non-interacting resonance gas which has an infinite radius
of convergence. At present, however, the large errors on $\rho_n (T)$ 
prohibit to draw a firm 
conclusion on the asymptotic behavior of $\rho(T)$. 
 
Similar caution is requested when interpreting the large fluctuations in
the quark number densities signaled by the rapid rise of the quark number 
susceptibility (see Fig.~5). As mentioned above we expect $\chi_q$ 
to diverge at a $2^{nd}$ order transition point. 
The rapid increase of $\chi_q$ is, however, partly due to the
rapid rise of $n_q$ itself as well as due to the rapid increase in the
pressure for $T\sim T_0$. A quantity reflecting the relative
magnitude of fluctuations is given by the ratio,   
\begin{equation}
\frac{\Delta p}{T^2 \chi_q} =
\frac{1}{2}\;  \biggl( {\mu_q \over T}\biggr)^2 { 
1 + \frac{c_4}{c_2} \left( {\mu_q \over T}\right)^2 
+ \frac{c_6}{c_2} \left( {\mu_q \over T}\right)^4  \over
1 + 6 \frac{c_4}{c_2} \left( {\mu_q \over T}\right)^2 
+ 15 \frac{c_6}{c_2} \left( {\mu_q \over T}\right)^4} \quad .
\label{ratio} 
\end{equation}
This quantity only depends on ratios of Taylor expansion coefficients,
and should vanish at a $2^{nd}$ order phase transition point. Its
temperature dependence is
shown in Fig.~\ref{fig:ratios_coefficients}(right). As 
can be seen the ratio rises rapidly across $T_0$ and approaches the ideal gas 
value. It, however, does not show any sign for a drop in the vicinity of $T_0$
that could be taken as evidence for the existence of a second order
transition.
 
\section{Resonance gas versus lattice results}

\begin{figure}
\begin{center}
\epsfig{file=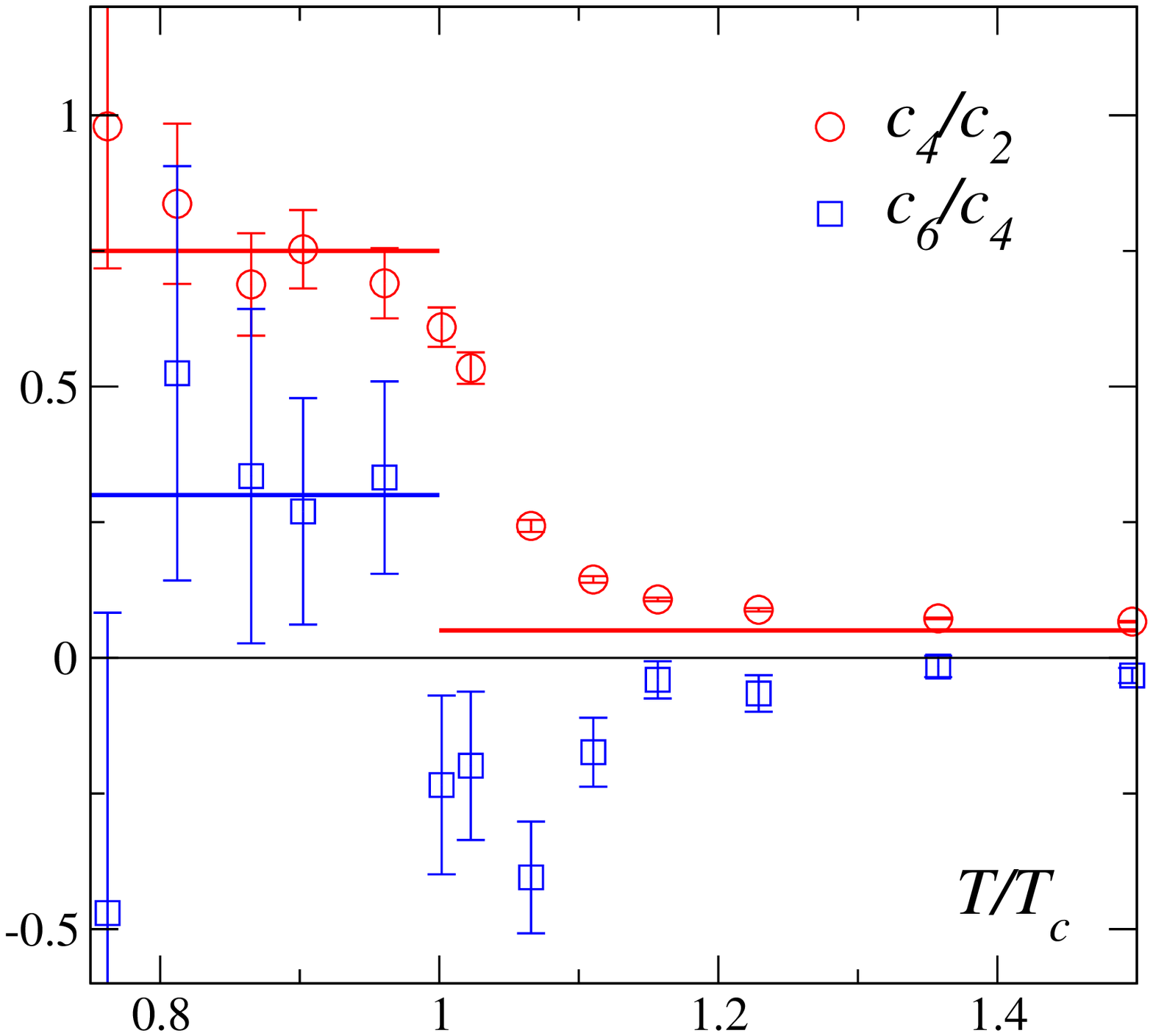,width=5.9cm}\hspace*{0.5cm}
\epsfig{file=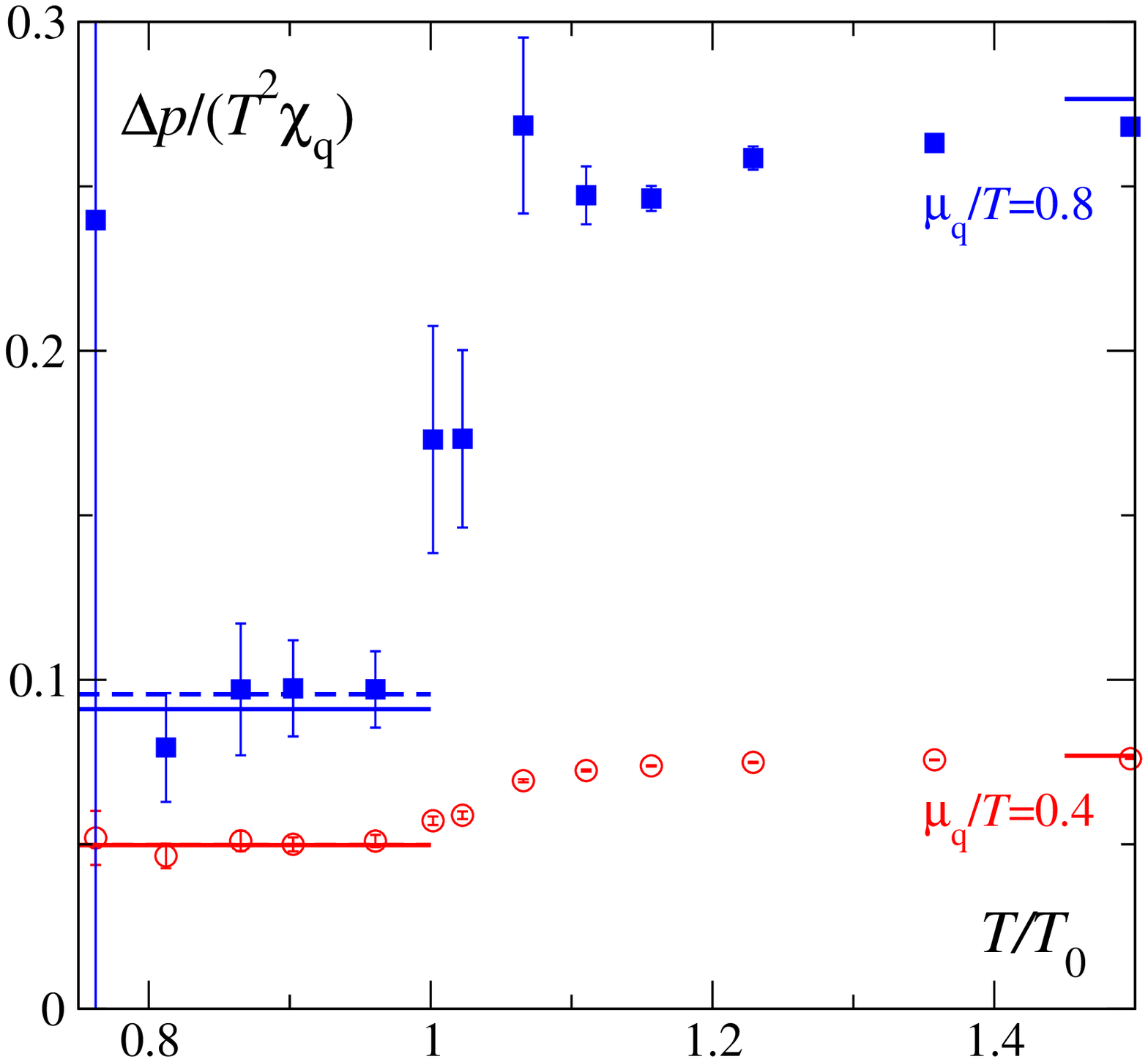,width=5.9cm}
\end{center}
\caption{\label{fig:ratios_coefficients} Ratios of subsequent expansion 
coefficients of the pressure (left) and the dimensionless ratio 
$\Delta p/T^2 \chi_q$ for two values of the chemical potential $\mu_q /T$ (right).
Horizontal lines show the expected results for a hadronic resonance gas and an
ideal quark-antiquark gas below and above $T_0$, respectively.
}
\end{figure}
Aside from the obviously rapid change of bulk thermodynamic observables
at temperatures close to $T_0$ it is a 
striking feature of the Taylor expansion that ratios of the expansion
coefficients are, to a first approximation, temperature independent below $T_0$ 
and distinctively different from the ratios above $T_0$. 
While for $T>T_0$ the ratios rapidly approach the corresponding
ideal gas values their low temperature behavior can be
understood in terms of the properties of a hadronic resonance gas \cite{redlich}. 
For $T\; \lsim\; T_0$ the baryonic contribution to the pressure
arises entirely from hadrons, which have masses significantly larger than
the relevant temperatures, {\it i.e.} $m_{baryon} \ge m_P \; \gsim\; 5 T_0$.
Their contribution may be approximated by a non-interacting resonance
gas obeying Boltzmann statistics. The pressure is then given by, 
\begin{equation}
\left( \frac{\Delta p}{T^4} \right)_{\rm res.~gas} \equiv 
f_B(T)\left( \cosh (\mu_B /T) -1 \right) \quad ,
\label{resgas}
\end{equation}
where $f_B(T)$ contains all the information on the resonance mass spectrum. 
Within this approximation the ratio of Taylor expansion coefficients is
given by $(c_{n+2}/c_n)_{\rm res.~gas} = 9/((n+2)(n+1))$ and thus
is indeed temperature independent. 
Moreover, the ratios are not affected by details of the hadronic
mass spectrum and, in particular, are not affected by the quark mass
dependence of the hadron spectrum. A straightforward comparison with
lattice calculations which generally are performed at unphysically large 
quark mass values thus is justified. Similarly, the ratio $\Delta p/T^2\chi_q
= (1-1/\cosh (3 \mu_q/T))/9$ is independent of the resonance spectrum $f_B(T)$
and does not vary with temperature for fixed $\mu_B/T$. 
These estimates are shown as straight lines in Fig.~\ref{fig:ratios_coefficients} 
for temperatures below $T_0$. Of course, the simple form of the $\mu_B$-dependence
in the resonance gas, Eq.~\ref{resgas}, does not lead to any critical behavior
in its own. At all temperatures the quark number susceptibility will
rise with increasing $\mu_q/T$ and for fixed $\mu_q/T$ it rapidly rises
with temperature. Moreover, as in the resonance gas model the dependence 
of e.g. the pressure on $\mu_q/T$ only appears through the factor $\cosh(3\mu_q /T)$
the radius of convergence of a Taylor expansion, of course, is infinite. 
{\it i.e.} 
$\displaystyle{\lim_{n \rightarrow \infty} (c_{n+2}/(c_{n})_{\rm res.~gas} =0}$. 
Any unambiguous evidence for the existence of a chiral critical point in the 
QCD phase diagram thus should also show up as clear deviation from resonance
gas behavior. 

\begin{figure}
\begin{center}
\epsfig{file=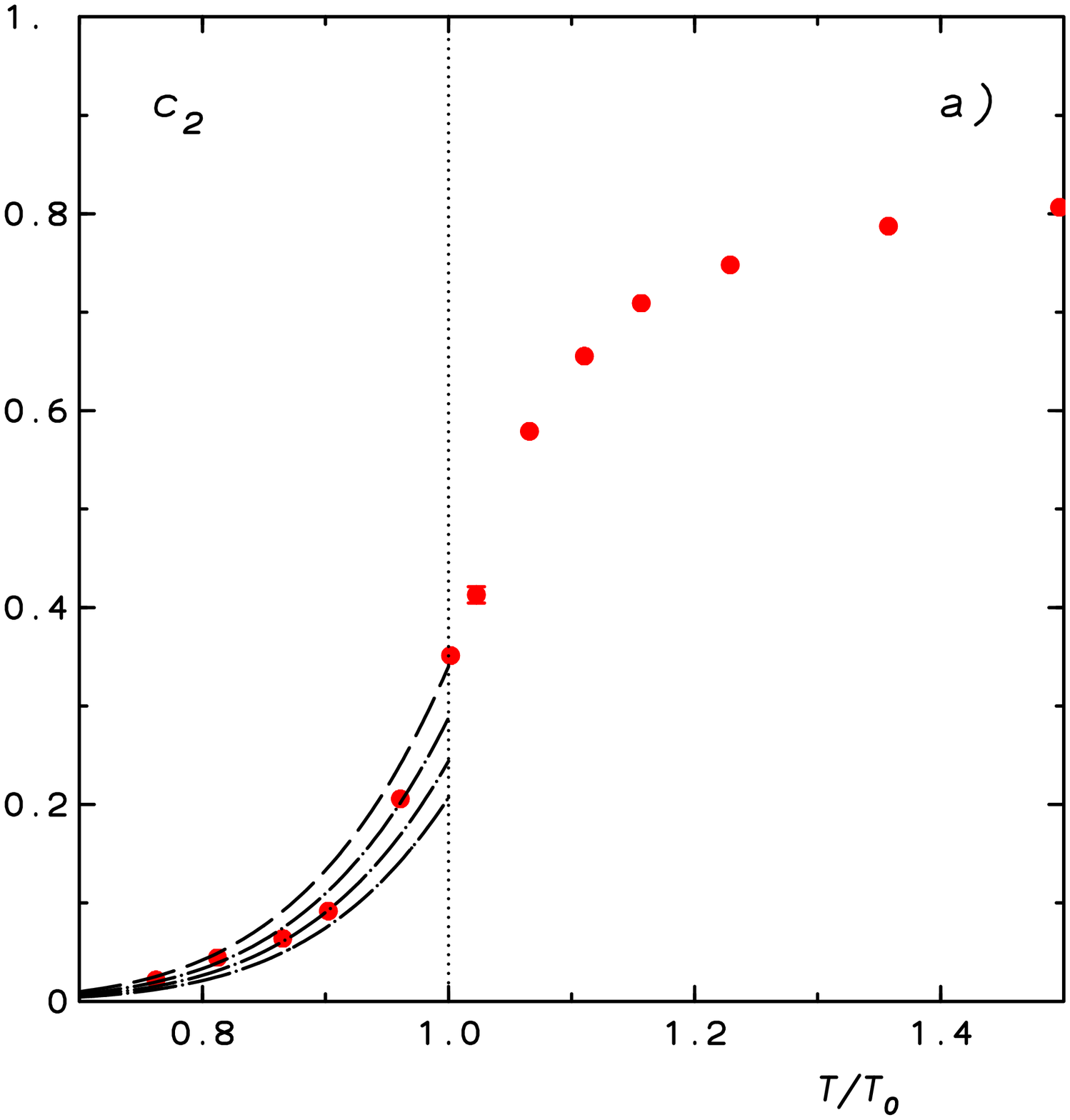,angle=180,width=5.5cm}\hspace*{0.5cm}
\epsfig{file=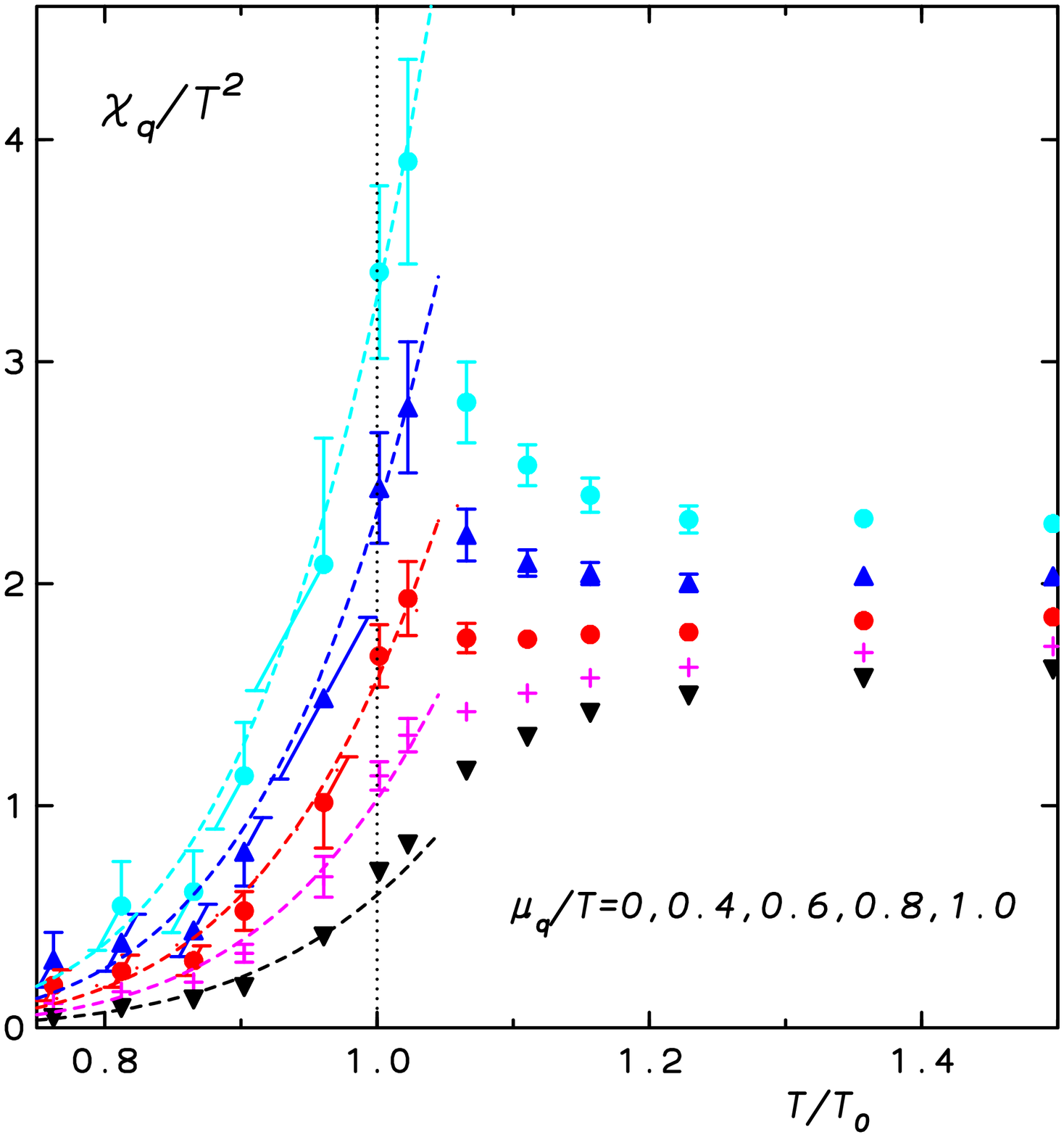,angle=180,width=5.5cm}
\end{center}
\vspace*{-1.0cm}
\caption{\label{fig:spectrum} Comparison of the resonance gas model
prediction for the
temperature dependence of the coefficient $c_2(T)$ with lattice results
(left) and the resulting dependence of $\chi_q /T^2$ on temperature (right)
for various values of the chemical potential. Shown are results from 
Ref.~\cite{redlich} where the resonance gas is compared to lattice calculations
up to ${\cal O}(\mu_q^4)$ only. The different curves shown in the left
hand figure correspond to $A= 0.9,~1.0,~1.1$ and 1.2 (top to bottom).
}
\end{figure}

A direct comparison of resonance gas model calculations for thermodynamic 
observables like the pressure or susceptibility with lattice calculations is
somewhat less stringent as it is necessary to adjust the resonance spectrum
to the still unphysical conditions that, at present, could be realized on the 
lattice. The
Taylor series discussed in section~3 is based on lattice calculations 
performed
with quark masses that correspond to a pseudo-scalar (pion) mass of about 
770~MeV. In this case the mass of the lightest baryonic state is about
twice as large as the experimental value for the nucleon mass.  This quark
mass dependence of the baryon spectrum, $m_H (m_{PS})$, has been modeled using 
the ansatz
$m_H (m_{PS})/m_H = 1 + A (m_{PS}/m_H)^2$ \cite{redlich}, where $m_H$ is
the experimental value of baryon resonances and $m_{PS}$ is the pseudo-scalar
meson mass used in the lattice calculations. A comparison with the first
expansion coefficient which only depends on the baryonic part of the 
spectrum\footnote{The leading term, $c_0(T)$, receives contributions from
the mesonic as well as the baryonic part of the spectrum.}, $c_2(T)= 9 f_B(T)/2$,
shows that the temperature dependence for $T\lsim T_0$ is well reproduced 
with $A\simeq 1$. With this also the temperature dependence of the 
quark number susceptibility is well described by the resonance gas in the
low temperature phase (see Fig.~\ref{fig:spectrum}). This comparison can
be extended to the meson sector contributing to $p/T^4$ at $\mu_B=0$ or 
to other quantum number channels controlled by other combinations of
the quark chemical potentials, for instance the iso-vector channel with
$\mu_I \equiv \mu_u -\mu_d$. Also in these cases lattice results for 
thermodynamic observables at low temperature are
well described by a hadronic resonance gas model \cite{redlich_sqm}.

\section{Conclusions}

We have shown that the thermodynamics of QCD at small values of the baryon
chemical potential can well be studied in terms of a Taylor expansion 
of the logarithm of the QCD partition function. In fact,
a $6^{th}$ order expansion seems to be quite sufficient for a reliable determination
of the change in pressure as well as a determination of the temperature
dependence of the net baryon number density for $\mu_B\lsim 3T$. 
The good agreement between lattice results in the low temperature hadronic phase
and resonance gas models is on the one hand conceptually appealing
as the resonance gas model has also been so successful in the 
description of experimental data on observed particle ratios in heavy
ion collisions \cite{pbm}. On the other
hand it shows that higher order calculations with high accuracy 
and/or calculations at smaller
quark masses are needed in order to find deviations from resonance gas
behavior and to deduce evidence for the existence of a chiral critical point in 
the QCD phase diagram.  

\section*{Acknowledgments}
This work has been supported by the German ministry for education and 
research (BmBF) under contract no. 06BI106 and grant of the 
{\it Gesellschaft f\"ur Schwerionenforschung} (GSI) under contract
BI/KAR.

\end{document}